\documentclass[preprint]{aastex6}

\begin{document}                                                                     

\title{X-rays in Cepheids:
{\it XMM-Newton} Observations of $\eta$ Aql 
\footnote{Based on observations obtained with  {\it XMM--Newton}, an 
ESA science mission with instruments and contributions directly 
funded by ESA Member States and the USA (NASA)}
} 


\author{Nancy Remage Evans}
\affil{Smithsonian Astrophysical Observatory,
MS 4, 60 Garden St., Cambridge, MA 02138; nevans@cfa.harvard.edu}

\author{Ignazio Pillitteri}
 \affil{INAF-Osservatorio di Palermo, Piazza del Parlamento 1,I-90134 Palermo, Italy}

\author{Pierre Kervella}
\affil{LESIA, Observatoire de Paris, Universit\'e PSL, CNRS, Sorbonne Universit\'e, Universit\'e de Paris, 5 Place Jules Janssen, 92195 Meudon, France}

\author{Scott  Engle}
\affil{Department of Astronomy and Astrophysics, Villanova University, 800 Lancaster Ave., 
Villanova, PA, 19085, USA}

\author{Edward Guinan}
\affil{Department of Astronomy and Astrophysics, Villanova University, 800 Lancaster Ave., 
Villanova, PA, 19085, USA}

\author{H. Moritz G\"unther}
\affil{Massachusetts Institute of Technology, Kavli Institute for Astrophysics and
Space Research, 77 Massachusetts Ave, NE83-569, Cambridge MA 02139, USA}

\author{Scott Wolk}
 \affil{Smithsonian Astrophysical Observatory,
MS 4, 60 Garden St., Cambridge, MA 02138}

\author{Hilding Neilson}
 \affil{Department of Astronomy and Astrophysics, University of Toronto, 
50 St. George Street, Toronto, ON, Canada M5S3H4}

\author{Massimo Marengo}
 \affil{Department of Physics and Astronomy, Iowa State University, Ames, IA, 50011,
 USA}

\author{Lynn D. Matthews}
 \affil{Massachusetts Institute of Technology, Haystack Observatory, 99 Millstone Rd., 
Westford, MA 01886, USA}

\author{Sofia Moschou}
 \affil{Smithsonian Astrophysical Observatory,
MS 4, 60 Garden St., Cambridge, MA 02138}

\author{Jeremy J. Drake}
 \affil{Smithsonian Astrophysical Observatory,
MS 4, 60 Garden St., Cambridge, MA 02138}

\author{Joyce A. Guzik} 
 \affil{Los Alamos National Laboratory, Box 1663,  MS T-082
Los Alamos NM 87545-2345}

\author{Alexandre Gallenne}
\affil{Nicolaus Copernicus Astronomical Centre, Polish Academy of Sciences, Bartycka 18, 00-716 Warszawa, Poland}
\affil{Departamento de Astronomia, Universidad de Concepcion, Casilla160-C, Concepcion, Chile}
\affil{Unidad Mixta Internacional Franco-Chilena de Astronomia (CNRS UMI 3386), Departamento de Astronomia, Universidad de Chile, Camino El Observatorio 1515, Las Condes, Santiago, Chile}

\author{Antoine M\'erand}
 \affil{European Southern Observatory, Karl-Schwarzschild-Str. 2, 85748 Garching, Germany}

\author{Vincent Hocd\'e}
 \affil{Universit\'e Cote d'Azur, Observatoire de la Cote d'Azur, CNRS, Laboratoire Lagrange, Boulevard de 
l'Observatoire, CS 34229 06304 Nice Cedex 4, France}
\affil{Nicolaus Copernicus Astronomical Centre, Polish Academy of Sciences, Bartycka 18, 00-716 Warszawa, Poland}



\begin{abstract}
 
 X-ray bursts have recently been discovered in
the Cepheids $\delta$ Cep and $\beta$ Dor
modulated by the pulsation cycle.
We have obtained an 
observation of the Cepheid $\eta$ Aql with the XMM-Newton
satellite at the phase of maximum radius,
the phase at which there is a burst of X-rays in $\delta$
Cep.  No X-rays were seen from the Cepheid $\eta$ Aql at this phase, and
the implications for Cepheid upper atmospheres are discussed.
We have also used the combination of X-ray sources and Gaia
and 2MASS data to search for a possible  grouping around 
the young intermediate mass Cepheid.  No indication of 
such a group was found.


\end{abstract}




\section{Introduction}

 Cepheids are particularly important because the Leavitt (Period-Luminosity)
Law provides distances which are the first step in the extragalactic distance
ladder.  Despite this, Cepheids are still poorly understood.
X-ray observations are a valuable tool for understanding Cepheids since
they provide insights into the physics of the upper atmosphere.
 Their properties as members of multiple systems also provide 
information about star formation and evolution for intermediate mass stars.

 \subsection{X-rays in Cepheids}  
 Recently  XMM-Newton observations of the Cepheid archetype
 $\delta$ Cep itself have found an increase in X-rays
in a very limited pulsation phase range (Fig. 1 in Engle,
et al. 2017). For most of the pulsation cycle, X-ray 
flux is modest ($\log{L_X}$  = 28.5-29.1 erg sec$^{-1}$),
appropriate for a coronal supergiant. However, near
the maximum radius (phase 0.5) the X-ray flux rises
rapidly, and then falls rapidly 0.10 later in phase.
Maximum luminosity  ($\log {L_x}$  = 29.23 erg sec$^{-1}$)
is four times minimum luminosity.
The phase in the pulsation cycle at which this occurs
is particularly surprising. At phases just after the
minimum radius (the “piston phase”) when the atmosphere 
is given a “push” by the envelope pulsation
cycle, many disturbances are seen in the photosphere
and chromosphere: ultraviolet lines in emission (see 
Fig. 1 in Engle, et al.), 
and increased turbulence.  Near maximum radius,
however, such signs of disturbance are absent, and, in fact, the 
spectra of Cepheids are indistinguishable from nonvariable stars.

The pattern of increased X-ray emission 
at maximum radius has been  confirmed in 
two cycles of  $\delta$  Cep 
(pulsation period of 5$^d$)
(Engle, et al. 2017).
  $\beta$ Dor, on the other hand, has a pulsation period of 10$^d$ 
where the light curve is distorted at maximum light (the standard fiducial 
for calculating phases) 
by the coincidence of primary and secondary pulsation maxima. 
 If instead we use 
the appearance of chromospheric emission lines  to
mark the phase of minimum radius (as  for $\delta$ Cep), 
the resulting phase of X-ray emission (after minimum radius)
is the same in both $\delta$ Cep and $\beta$ Dor, that is 
just after maximum radius,  as shown   
in Fig. 6 in Evans, et al. (2020a).

The occurrence of increased  X-rays  at a specific phase of the 
pulsation cycle ties the phenomenon to  pulsation. 
The link to pulsation has long been suspected 
in possible mass loss scenarios, and X-rays may now tie upper
atmosphere activity to pulsation.  Since the 
photosphere and chromosphere are quiescent at this phase
(maximum radius), the 
reasonable explanation  is that the 
disturbance results as  pulsation expansion progresses to the outer 
atmosphere. Based on this interpretation, a reasonable velocity (35 km s$^{-1}$) and 
the time between minimum and maximum radius  
 indicate that the X-ray activity occurs at about 0.3 R$_{Cep}$ above
the photosphere.

There are two possible causes of the X-ray bursts.
 1. One possibility is that {\it the pulsation cycle itself generates a shock wave}.
Velocities  seen in Cepheid photospheric pulsation are typically
35 km sec$^{-1}$, which would have to accelerate outward (e.g. due to a pressure
gradient)
to explain the X-rays.  2. The other possibility is a 
{\it coronal reconnection event} (flare)  such as the ones that frequently 
occur on the sun.   A magnetic field is 
required for this. However, the occurrence of X-rays at all 
phases of Cepheids  (albeit at a low level: Engle, et al. 2015)
 is  a strong indication of a magnetic field.
We have  undertaken  a theoretical 
modeling program  to explore these
possibilities. 
Moschou et al. (2020) have  modeled the  pulsation driven shock
corona of  $\delta$ Cep using the code PLUTO (Mignone 2014) for a pure hydrodynamic (HD)
 setup. Models include
atmospheric stratification in spherical geometry and a simple sinusoidal
driver for the pulsation at the bottom of the stellar corona.
The models indicate
that under specific conditions shocks are able to reproduce a phase
dependent X-ray luminosity enhancement for pulsation driven outflow.

 We have embarked on a program of X-ray observations of Cepheids
to explore the parameter dependence of X-rays 
(Cepheid Outer Atmospheres: X-rays [COAX]).  The observation of $\eta$ Aql discussed here
is part of this program.

\subsection {CSEs} The pattern of X-ray observations may provide a clue to 
upper atmosphere phenomena in the pulsating atmosphere of Cepheids.  Another 
recently observed phenomena which may be related is circumstellar 
envelopes (CSEs).  
 Excess infrared (IR) emission around Cepheids was first identified in interferometric
results, as summarized by Gallenne, et al (2021).  The first cases were l Car (Kervella, et al.
2006), Polaris and $\delta$ Cep (M\'erand, et al. 2006).
It was subsequently 
found in IR photometry, including from the Spitzer satellite.  
Recent discussions are found in Gallenne, et al. and Groenewegen (2020), 
 Marengo, et al. (2010 a,b)
Barmby et al. (2011), Scowcroft et al. (2016) and Schmidt (2015). 
A comprehensive  fit of photometry, velocities and angular diameters 
is provided by the  
SpectroPhoto-Interferometry of Pulsating Stars (SPIPS) from  M\'erand, et al (2015),
which included IR excesses.
The IR excess is small (mean values from Gallenne, et al.:
0.09 $\pm$ 0.03 mag at 2.2 $\mu$m, 
0.14 $\pm$ 0.04 mag at 10 $\mu$m).  

The source of the emission is uncertain.  The most recent discussion (Hocd\'e et al. 2020a)
has examined energy distributions including Spitzer spectra.  They find hot or cold dust 
cannot explain the spectral distribution, however free-free emission from a thin shell of 
ionized gas  at about 0.15  R$_{Cep}$ 
does match the emission.  Hocd\'e et al. (2020b) used lines of H$\alpha$ and 
the IR Ca triplet to study the chromosphere.  For Cepheids with periods longer than 10$^d$, 
they find that the thickness of the chromosphere is about 50\% of the radius of the star.  
This could possibly contain a hot inner chromosphere where the lines are found and an 
cold outer chromosphere seen in interferometry.  
Furthermore,
for most of the long period Cepheids they find a motionless H$\alpha$ absorption feature at 
the stellar rest frame, which could come from the outer CSE.  This is similar to the 
stationary absorption line found in the Mg II profiles by B\"ohm-Vitense and Love (1994).

CSEs are related to two aspects of Cepheids.  First IR flux must be taken into account for the 
most accurate application of the Leavitt law.  Second  pulsation may cause mass-loss, even at
a very low level (see Neilson, et al. 2012 for a recent summary).  This would affect the evolution
of the Cepheid and the interpretation of evolutionary tracks.  

  \subsection{Outer Atmosphere}
To emphasize the important niche that the X-ray observations play in 
understanding the upper atmosphere of Cepheids, 
we summarize the phenomena and what can be inferred about stratification
between them, particularly as they are seen in $\delta$ Cep.

 1. In the inner region  of a Cepheid atmosphere 
we have many diagnostics for the
pulsation related disturbances in the  photosphere and chromosphere at 
minimum radius.   

 2. The layer related to 
 X-ray increase is above this and  does not participate
in the disturbances  at minimum radius, nor do the photosphere
or chromosphere show any disturbance at the time of X-ray maximum (maximum radius).

 3. The CSEs are are identified by IR emission.   
At present  we are accumulating diagnostics for the outer atmosphere
CSEs and X-ray region, but it is not clear how they are related.    

   4. Beyond the CSE and X-rays, $\eta$ Aql has two companions (see below) which might sculpt 
any mass-loss flow.

 5. In $\delta$  Cep itself   there is a  spectacular 
shell (bow shock) which appears to  surround the
Cepheid which could be created by  a mass loss wind 
from the Cepheid interacting with the ISM (Marengo et al. 2010b).  
However, this has not been seen in   $\eta$ Aql.  

\subsection{$\eta$ Aql} 

The target of the  XMM-Newton observations, $\eta$ Aql, is one of the brightest
Cepheids, and has been extensively observed.
It has a pulsation 
period of 7.18$^d$, an E(B-V) of 0.12 mag, and a distance of 273 pc (Evans, et al. 2016
 based on the HST parallax scale of Benedict, et al. 2007).

Available data for 
$\eta$ Aql are assembled in M\'erand, et al (2015) in the demonstration of the 
SPIPS program.  These include interferometry
from Kervella, et al. (2004) and Lane et al. (2002).  M\'erand et al. found an excess of 
0.018 $\pm$ 0.002 mag in the K band and 0.016 $\pm$ 0.003 mag in the H band.
This analysis was recently redone by Gallenne, et al (2021) adding new observations 
from the VISIR instrument at the Very Large Telescope.  They find IR excesses
increasing with wavelength from
0.077 $\pm$ 0.005 at {\it K} 
to  0.20  $\pm$ 0.01  mag at 25 $\mu$m.

 $\eta$ Aql has two companions. 
 The closest is a  B9.8 V star (Evans et al. 2013). The orbit is not yet known, 
but as post-red giant stars, all known Cepheid binaries 
have separations of at least 1 AU.   
  The more distant companion is an early F star 
0.66$"$ or 180 AU from the Cepheid (Gallenne, et al. 2014).
 Stars in the spectral range late B through early F do not in 
general produce X-rays, so 
neither of these companions is expected to produce 
X-rays.

The purpose of of the XMM-Newton observation was to determine whether there is an X-ray burst at maximum radius 
as is seen for $\delta$ Cep.

\subsection{Possible companions?}

Reasonably massive stars like Cepheids (typically 5M$_\odot$) are frequently found in binary or 
multiple systems (e.g. Evans. et al. 2020b).  In addition a number are known in open clusters, which 
have been an important source of calibrators for the Leavitt Law.   Anderson, Eyer, and Mowlavi (2013) 
provide a recent assessment of Cepheid  membership in clusters.  
Gaia data make it possible to investigate whether there 
are any  associations with low densities between recognized clusters and multiple systems.  
X-ray observations
add an important element.
Low mass stars at the age of Cepheids are X-ray active.  This means X-ray 
observations are particularly useful at distinguishing low mass stars related to Cepheids from the 
older field star population.  This is demonstrated, for instance, for the Cepheid S Mus (Evans, et al.
2014).  We have used the XMM-Newton observation of $\eta$ Aql combined with 2MASS and Gaia data to 
search for any low mass X-ray active stars in the field.

The following discussion of the XMM-Newton observations of $\eta$ Aql
has the following sections: the observation and data analysis, 
the X-rays at the position of the Cepheid, identification of counterparts to X-ray sources in 2MASS and Gaia,
and discussion and conclusions.

\section{Observation and Data Analysis}


\begin{figure}
\plotone{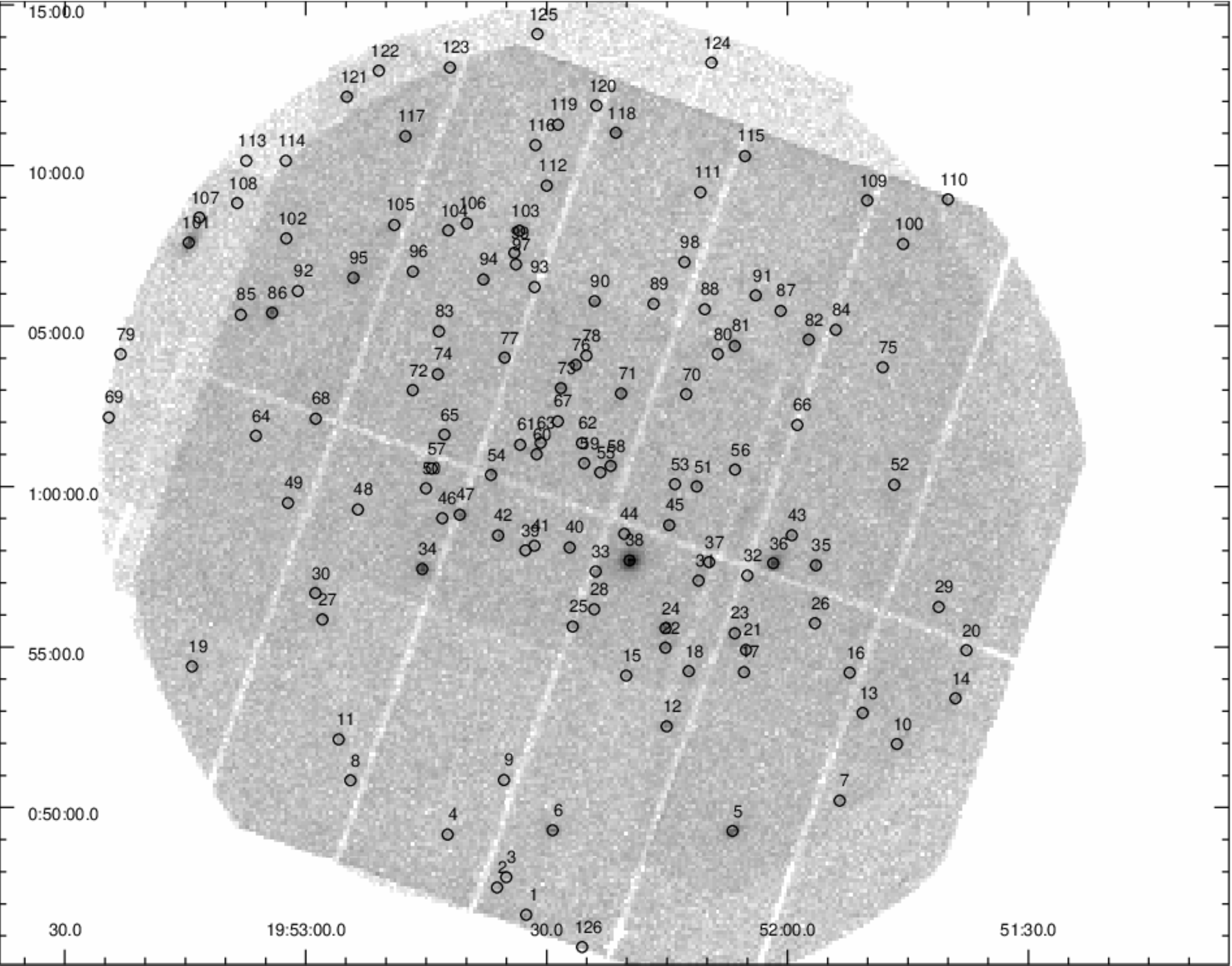}
\caption{The XMM-Newton image of the field around $\eta$ Aql.
The image is the combined  PN, MOS1 and MOS2 data. 
 The numbers   show the locations of the 
sources in Table 1.
\label{img}}
\end{figure}

\begin{figure}
\plotone{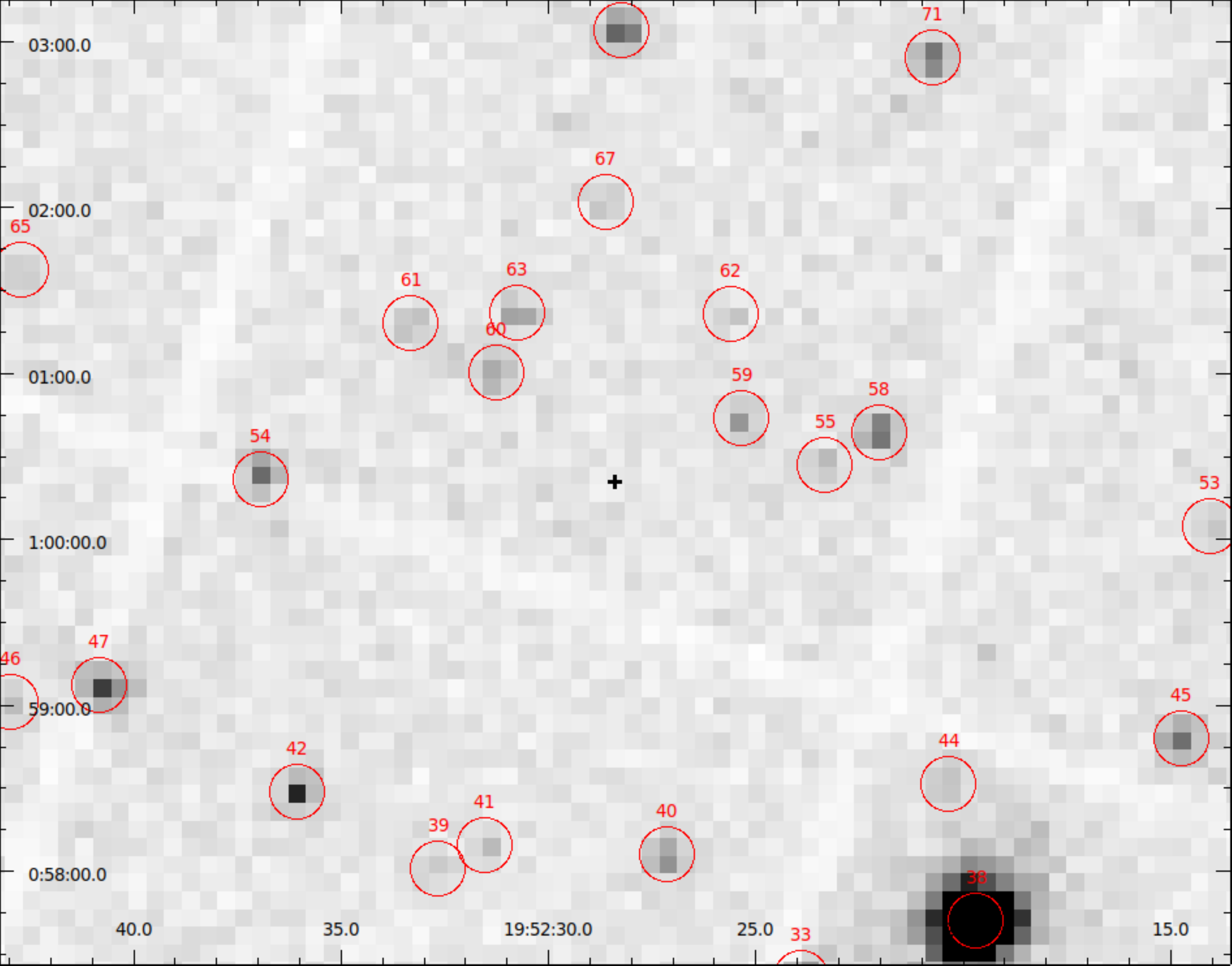}
\caption{A zoom of the center of the image in Fig.~\ref{img} with the position of the 
Cepheid marked with +.
\label{zoom}}
\end{figure}

$\eta$ Aql was observed by {\it XMM-Newton} for 75 ksec on May 12, 2019 
from JD 2,458,615.584 to  2,458,616.452 (OBSID 0840510201).  
Data analysis was carried out using standard data reduction 
tasks in SAS  (Scientific Analysis Subsystem) software version 17.0
 as in Pillitteri, et al. (2013).  This involved a reduction 
starting from the ODFs (Observation Data Files) of the observation, filtering the events 
according to their grades
and screening out bad pixels. Only events between 0.3 and 8.0 keV 
were used.  Using the recipe given in the SAS 
guidelines 
the reduction was restricted to good time intervals and low 
background periods,
 based on the light curve of the events above 10 keV.
Fig.~\ref{img} shows the detected sources.  
The X-ray positions have a median 
uncertainty in position of 2.9$\arcsec$ with a range of  2.3$\arcsec$ to 4.0$\arcsec$
 (25\%-75\% range of the distribution).

A full list of the X-ray sources is provided in the electronic version, with a section in Table 1 below to 
illustrate form and content.  The columns show Id, RA, Dec, position error, distance  from the aim point,
 the significance of detection in units of standard deviation of the local background, the rate with error 
scaled to the sensitivity of MOS and the exposure time.
The exposure time is the sum of MOS and pn exposures.

\begin{deluxetable}{llccccccc}
\tablecaption{ Sources in the {\it XMM-Newton}  Image around $\eta$ Aql  \qquad\qquad \label{src}}
\tablewidth{0pt}
\tablehead{
 \colhead{} &  \colhead{RA}  & \colhead{Dec}  &  \colhead{Pos Err}  & \colhead{Offaxis}  & \colhead{Significance}
 &  \colhead{Rate}  & \colhead{$\pm$}  & \colhead{Exp Time}    \\
\colhead{} &  \colhead{(J2000) }  & \colhead{(J2000)} &  \colhead{ $"$ }
 & \colhead{$'$}  & \colhead{$\sigma_{bkg}$}
 &  \colhead{counts ks$^{-1}$}  & \colhead{counts ks$^{-1}$}  & \colhead{ks}    \\
}
\startdata
1 &  298.13559 &	0.77763 &  1.2 &	14.89 &	11.6 &	1 &	0.16 &	54.14 \\
2 & 298.15073 &	0.79184 &	2.9 &	14.23 &	5.5 &	0.61 &	0.15 &	58.38 \\
3 & 298.14593 &	0.79717 &	1.8 &	13.86 &	6.8 &	0.62 &	0.12 &	60.28 \\
 4 & 298.17638 &	0.81929 &	4.8 &	13.13 &	7.3 &	1.3 &	0.21 &	66.91 \\
5 & 298.02853 &	0.82115 &	2.6 &	12.72 &	35.8 &	9.3 &	0.45 &	73.99 \\
 6 & 298.12188 &	0.82154 &	2.9 &	12.16 &	24.6 &	5.7 &	0.35 &	69.57 \\
 7 & 297.97294 &	0.83689 &	3.1 &	13.33 &	10.5 &	1.8 &	0.26 &	53.65 \\
\enddata

\end{deluxetable}

\section{The Cepheid}

The observation covered phases 0.406 to 0.527
using the ephemeris from Engle (2015) for the epoch E:
\vskip .1truein
\noindent 2,455,856.689 + 7.177025 $\times$ E 
\vskip .1truein
\noindent This covers the phase range of the X-ray burst in the 5$^d$ Cepheid, $\delta$ Cep.


The Cepheid $\eta$ Aql was not detected.  Fig.~\ref{zoom} shows a zoom of the center of the XMM-Newton image with 
the location of the Cepheid marked with +.
The upper limit on the X-ray luminosity 
from this was estimated as follows. 
In a region of 30" around the optical position of the star there are 
528$\pm$ 23 counts in PN energy band 0.3-8.0 keV. The region should contain
about 80\% of the PSF. Using 3 $\sigma$ from the 23 counts, and correcting
for the 80\% factor results in 86 counts in 75 ks. 
Using  Portable Interactive Multi-Mission Simulator
(PIMMS) with an APEC component of 0.5 keV and
 $N_\mathrm H = 7\times10^{20}$ cm$^{-2}$
the upper  limit of the  flux is
 2.01 $\times$ 10$^{-15}$  ergs s$^{-1}$ cm$^{-2}$. 
 The upper limit  results an upper limit to the luminosity of 
$L_X \le 1.8\times10^{28}$ erg/s using a 
distance of 273 pc, which is lower than the luminosity 
$\delta$ Cep in its quiescent phases (3.2 to 12$\times$ 10$^{28}$ ergs s$^{-1}$   and certainly less than 
at its maximum phase, where the luminosity is 1.7 $\times$ 10$^{29}$ ergs s$^{-1}$. 

\begin{figure}
\plotone{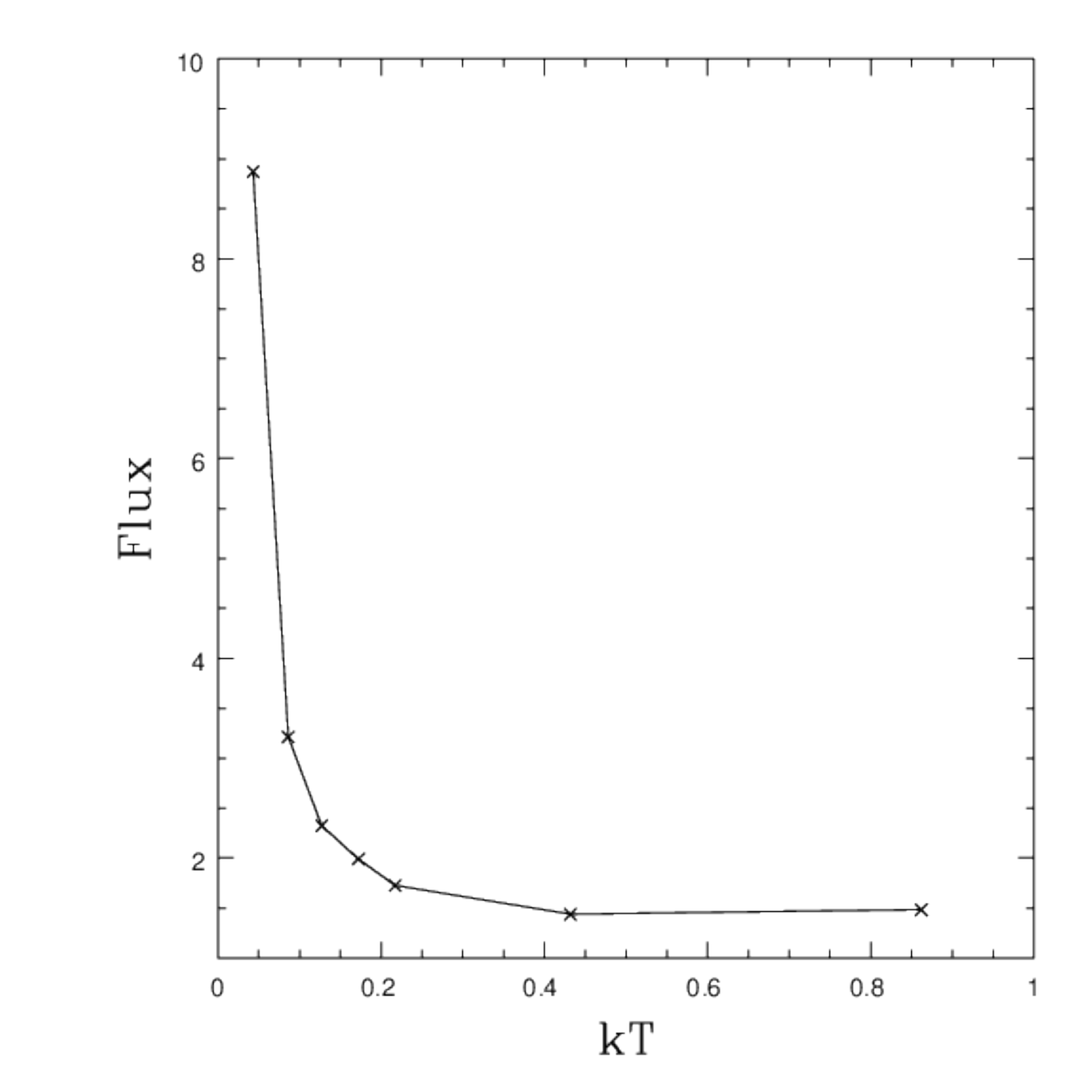}
\caption{The flux computed from the count rate for  the 0.3 to 1.0 keV band in 
Table 2}
for a range of temperatures (see text).  The 
temperature  kT is in keV; the flux is in 10$^{-15}$  ergs s$^{-1}$ cm$^{-2}$.  
\label{flux}
\end{figure}

We have investigated  to see how much this upper limit can be varied.  In particular, we have 
used narrower wavelength bands for comparison, as shown in Table~\ref{cul}.  Since Cepheids and
other supergiants have soft X-ray spectra, their flux is concentrated below 2 keV.  Thus, the second
row in   Table~\ref{cul} shows the upper limit to the count rate for the band between 0.3 and 1.0 keV.  
The bottom row in  Table~\ref{cul} shows the much narrower band between 0.45 and 0.67 keV which contains
the O VII triplet.  To interpret the count rate we provide the example for the  0.3 and 1.0 keV band 
of the flux provided by  
PIMMS for a series of temperatures in Fig.~\ref{flux}.  
For the range of temperatures between 0.4 and 
0.8 keV, the flux is essentially the same, so the flux of  ~1.4 x 10$^{-15}$   ergs s$^{-1}$ cm$^{-2}$ in
Table~\ref{cul} is temperature independent. Luminosities are given in the final 
column of Table~\ref{cul}, 
which are further below the luminosities of $\delta$ Cep in either its quiescent or maximum state.

\begin{deluxetable}{llcc}
\tablecaption{$\eta$ Aql Upper Limits  \qquad\qquad \label{cul}}
\tablewidth{0pt}
\tablehead{
 \colhead{Energy} &  \colhead{Count Rate} & \colhead{Flux} &  \colhead{Lum}   \\
 keV     &    x 10$^3$  & ergs s$^{-1}$ cm $^{-2}$      &  ergs s$^{-1}$  \\ 
}
\startdata
0.3 -- 8.0  &  1.15  &  1.7 $\times$ 10$^{-15}$ & 1.5 $\times$  10$^{28}$  \\ 
0.3 -- 1.0   &   $\simeq$1.0   &  $\simeq$1.4 $\times$  10$^{-15}$   & 1.2 $\times$  10$^{28}$  \\
0.45 -- 0.67  &  0.5   &  0.7 $\times$ 10$^{-15}$   &  0.6 $\times$ 10$^{28}$ \\
\enddata
\end{deluxetable}

\section{Identification of Counterparts to X-Ray Sources}

\subsection{2MASS Sources}

\begin{figure}
\plotone{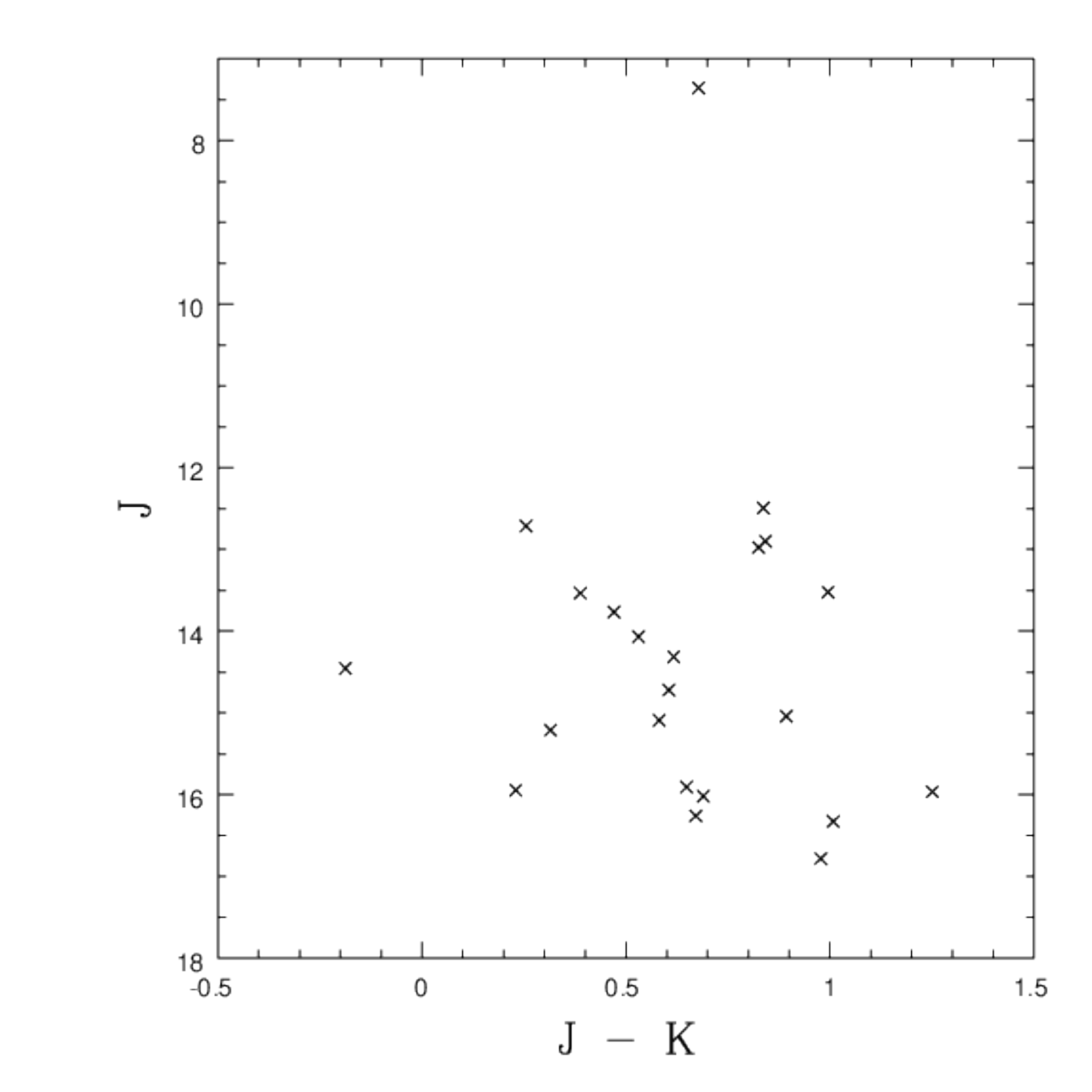}
\caption{The color-magnitude diagram for X-ray sources with 2MASS counterparts 
(in magnitudes) for 2MASS errors less than 0.1 mag.  
\label{2mass}}
\end{figure}

We have used the Two Micron All Sky Survey (2MASS) catalog  
(Cutri, et al. 2003) to find the near-IR counterparts to the X-ray sources. Thirty four   
have a 2MASS source within 5$"$.  The full list of the  2MASS counterparts
detected in the $\eta$ Aql field 
is provided in the electronic version, with a short list in  Table 3 to indicate the 
content of the table.  The columns show the XMM-Newton source ID, the RA, the Dec, the 
2MASS ID, the 2MASS magnitudes and errors (J, H, and K), the 2MASS quality indicators, 
and the separation between the X-ray source and the 2MASS source.  
Fig~\ref{2mass} shows the J--(J-K) color magnitude diagram for 
the 2MASS counterparts for sources where the errors on the photometry are less than 
0.1 mag.  An M0 star at the distance of $\eta$ Aql (273 pc) with the reddening of the 
Cepheid (E[B-V] = 0.12 mag   would have a J = 13.3 mag and
(J-K)$_0$ = 0.85 mag
using the main sequence calibration of Drilling and Landolt (2000), and the colors
calibrations of 
Bessell and Brett (1988) and Carpenter (2001).
Fig~\ref{2mass} shows that most of the 2MASS sources are fainter and/or redder 
than these values,
with only a few sources which could be at the same distance as the Cepheid.  Since 
the Cepheid is at a low galactic latitude (b = 13$^{\circ}$), it is likely that there are a number 
of young stars at larger distances and with larger reddenings, 
which would constitute the majority of the 2MASS 
X-ray sources.  
Furthermore, it is likely that  sources without 2MASS counterparts are 
background AGN.

  The bright star in Fig.~\ref{2mass} is HD 187900, 
which is a K2 star with V = 9.28 mag.

\begin{figure}
\plotone{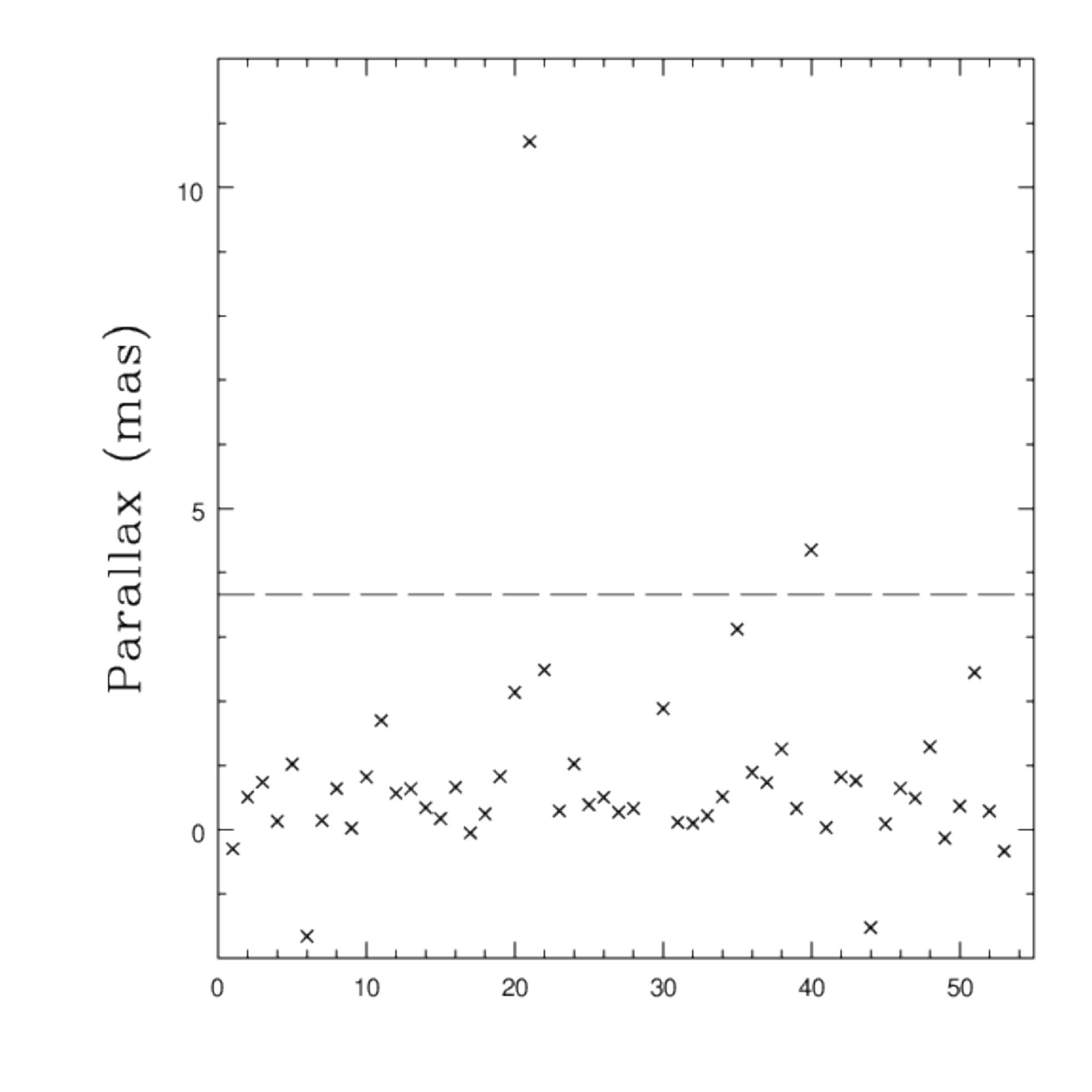}
\caption{{\it Gaia} parallaxes for possible matches with X-ray sources
plotted as a function of line number in Table 4.
The dashed line corresponds to a distance of 273 pc for $\eta$ Aql.
\label{x.gaia}}
\end{figure}

\begin{figure}
\plotone{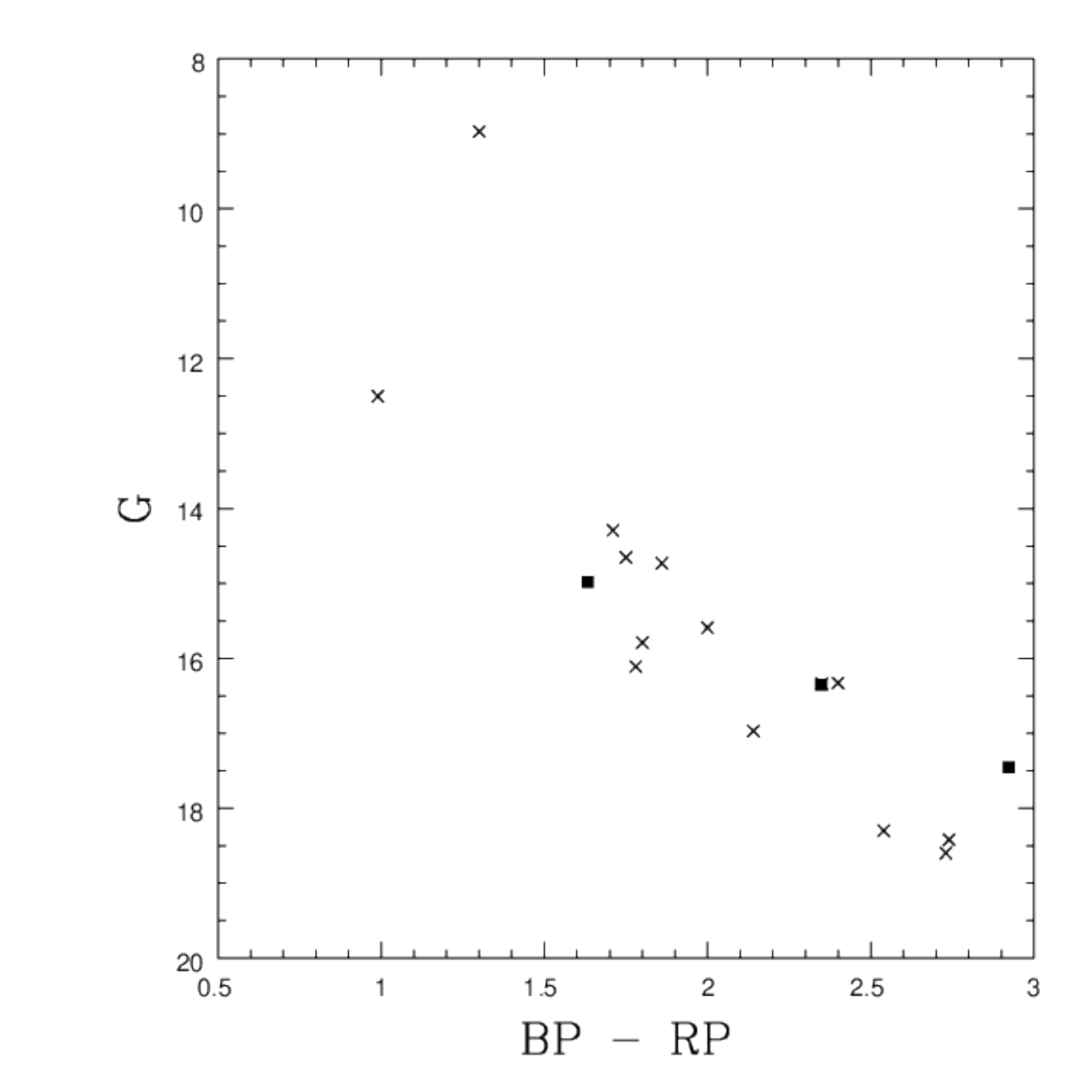}
\caption{The color-magnitude diagram for Gaia sources  
(in Gaia magnitudes) at approximately the distance of $\eta$ 
Aql.  The sources have parallaxes between 2.0 and 4.5 mas
and parallax errors less than 0.2 mas.  Gaia sources: x's; 
  sources with an X-ray
detection:  filled squares. 
\label{d.gaia}}
\end{figure}

\hfill\vfill
\eject



\begin{deluxetable}{llcccccccccc}
\tablecaption{2MASS Counterparts to   {\it XMM-Newton} Sources   \qquad\qquad \label{src}}
\tablewidth{0pt}
\tablehead{
 \colhead{} &  \colhead{RA}  & \colhead{Dec}  &  \colhead{2MASS ID}  & \colhead{J}  & \colhead{$\pm$}
 &  \colhead{H}  & \colhead{$\pm$}  & \colhead{K}   & \colhead{$\pm$} & \colhead{Qual}  & \colhead{Sep}   \\
\colhead{} &  \colhead{(J2000) }  & \colhead{(J2000)} &  \colhead{  }
 & \colhead{mag}  & \colhead{mag}
 &  \colhead{mag}  & \colhead{mag}  & \colhead{mag}  & \colhead{mag} & \colhead{}  
& \colhead{$"$}   \\
}
\startdata
 17 & 298.233307 & 0.868617  &  19525599+0052070 &  15.7430 &   0.0910  & 15.3750  &  0.1150 &  15.1300  &  0.1450 & ABB  &  1.3929 \\
 24 & 298.051544 & 0.903709  &  19521236+0054133 &  14.0700 &   0.0780  & 13.5690  &  0.1130 &  13.5390  &  0.0580 & ABA  &  1.0000 \\
 24 & 298.052582 & 0.903914  &  19521261+0054140 &  14.7210 &   0.0450  & 14.0680  &  0.0580 &  14.1160  &  0.0610 &  AAA &   1.0000 \\
 27 & 298.021759 & 0.914922  &  19520521+0054537 &  14.3140 &   0.0290  & 13.7460  &  0.0410 &  13.6970  &  0.0360 & AAA  &  1.2030 \\
\enddata

\end{deluxetable}

\subsection{ Gaia Sources}

The X-ray source list was also matched against the catalog produced by the 
 Gaia satellite.  (Gaia Collaboration 2016, 2018).
 The full list of possible matches is provided in the 
electronic version of this paper, with a few entries in Table~\ref{gsrc} to indicate the 
form and content. Of 126 X-ray sources, 54 (43\%) had at least one possible match with a 
 Gaia source.  The columns show the RA, the Dec, the separation of the XMM-Newton source and 
the Gaia source, the Gaia ID, the XMM-Newton source number, the parallax and error, and the Gaia
magnitudes and errors (G, BP and RP).  


\begin{table*}
\caption{ Gaia Counterparts to XMM-Newton Sources   \qquad\qquad \label{gsrc}}
\resizebox{0.95\textwidth}{!}{
  \begin{tabular}{llccccccccccc}\hline\hline
\tablewidth{0pt}
\hline
 RA  & Dec  &  Sep  &  Gaia ID  & X Src \#  & Px & $\pm$ &  G  & $\pm$  & BP   & $\pm$ & RP  & $\pm$   \\
 $^o$  & $^o$  & $"$  &          &          & mas & mas  &  mag  & mag  & mag   & mag & mag  & mag   \\
\hline
298.17575892226 &	0.81915965345 &	0.45 &	4240251955274546176 &	4 &	-0.2996 &	0.3579 &	19.3837	 & 0.0036 & 	19.6484 &	0.0707 &	18.7031 &	0.0462 \\
298.02863373075 &	0.82086525281 &	0.3 &	4240262920333547008 &	5 &	0.5077 &	1.1782 &	20.6907 &	0.0097 &	20.5658 &	0.1153	& 20.0888 &	0.1044 \\
298.12175960105 &	0.82134431695 &	0.22 &	4240252470670622720 &	6 &	0.7396 &	1.0209 &	20.5027 &	0.0093 &	20.471 &	0.1214 &	19.4289 &	0.0448 \\
297.97295268783 &	0.83683043657 &	0.06 &	4240263122188744704 &	7 &	0.1319 &	0.6752 &	20.1036 &	0.0095 &	20.3039 &	0.0777 &	19.5392 &	0.0558 \\
\hline
\end{tabular}
}
\end{table*}



The distribution of parallaxes is shown in Fig.~\ref{x.gaia}.  For comparison, the dashed line  indicates
the parallax of $\eta$ Aql, which corresponds to a distance of 273 pc.
There are other determinations of the 
distance (e.g. 296 pc from the SPIPS method [M\'erand, et al. 2015]).
Ultimately, it is anticipated the Leavitt law will 
 be improved by {\it Gaia},  however adjustments to Fig.~\ref{x.gaia} will be small. 
The possible matches largely correspond to distances of a kpc or more, well beyond the 
location of the Cepheid.  There are in fact only a couple of stars which might be at 
the Cepheid distance.   To identify possible structures in lines of sight in the
galactic plane, it is valuable to have  parallaxes and photometry, though in this 
case no structures were identified within 1 kpc.


As part of the exploration of the population in the direction of $\eta$ Aql, we made an 
additional comparison, using   a list of stars with {\it Gaia} parallaxes drawn up in
the same way as in Kervella, et al. (2019) to look for wide common proper motion 
companions of Cepheids.  
This 
search covered approximately the same area as the  XMM-Newton field.
We examined stars at approximately the distance of $\eta$ Aql
with parallaxes between 2.5 and 4.5 mas (corresponding to distances between 
approximately 220 and 400 pc), and  
further restricted the list to those with parallax errors less than 0.2 mas.  The resulting sample
is shown in Fig~\ref{d.gaia}.  As would be expected for this distance
 restricted sample the color magnitude 
diagram forms a sequence.  
For comparison, we show the  XMM-Newton sources which 
matched  Gaia sources.  Although there are three  XMM-Newton sources consistent with 
the cmd sequence, the Gaia  objects in general are not X-ray sources.  This suggests
 that while sharing a similar distance they are stars older than the Cepheid (50 Myr)
 and thus less X-ray luminous and not related to it.   (X-ray activity is 
typical in young low-mass stars at the flux which would be detected in this 
observation. However for a young high mass star, such as a 
Cepheid, X-rays occur  at this level
only in a restricted phase range, such as for $\delta$ Cep.  
  In $\eta$ Aql, the observation does not detect X-rays in this phase
range.)



\section{Discussion and Conclusions}

 In this paper we have found that 
the Cepheid $\eta$ Aql is not an X-ray source in the phase range at which an X-ray 
burst was seen in $\delta$ Cep.  
Furthermore, $\eta$ Aql has a stronger IR excess than $\delta$ Cep indicating a CSE.  Thus  
we do not seem to duplicate the pattern of X-ray occurrence and CSE seen in $\delta$ Cep.  

The Cepheids $\eta$ Aql and $\delta$ Cep have many characteristics which are similar:  
pulsation periods,  amplitudes and pulsation modes.
The pulsation 
periods imply that the masses and ages are similar.  The pulsation period of $\delta$ Cep is decreasing implying
that it is on the second crossing of the instability strip, in contrast to   $\eta$ Aql which has an 
increasing period implying that it is on the third crossing (Engle 2015).  This, however, is not a sign of 
different or abnormal evolution, rather just part of the standard evolutionary sequence.  
One parameter which might differentiate them is a magnetic field.  Little is 
known about the magnetic fields in Cepheids.  
Grunhut et al. (2011) find a Zeeman signature in $\eta$ Aql but not
in $\delta$ Cep.  What role this might play in X-ray production or CSEs  is not known. 

This raises the question of whether X-ray bursts are common or exceptional 
among Cepheids since we see them 
in two ($\delta$ Cep and $\beta$ Dor) but not in $\eta$ Aql. Alternately, 
is there anything about $\eta$ Aql
which  would make it exceptional?  The general properties of $\eta$ Aql
(period, amplitude pulsation mode) are all 
common among many other Cepheids.   
 An example of a property which might affect X-ray production and
be unique to  $\eta$ Aql is a companion which comes close to the Cepheid 
at periastron because of high eccentricity.  $\eta$ Aql does have a close
companion, but the orbit has not so far been determined.


The X-ray peaks from the theoretical simulation do not all occur 
  at the same phase. Among other things,
 the phasing depends on the number of terms in the Fourier representation of the pulsation 
wave.  This suggests another way the occurrence of an X-ray increase may depend on stellar 
parameters, and may differ from star to star.

In summary, the pattern of an X-ray burst in $\delta$ Cep is not repeated in 
$\eta$ Aql, despite the fact that they are similar in many physical parameters.
The links between upper atmosphere phenomena are apparently not simple.







Stars in the region of $\eta$ Aql 
that are young enough to be associated with the Cepheid are expected to be
X-ray sources.  
 Fig~\ref{d.gaia} shows sources in the XMM-Newton field-of-view  
 which have reasonably accurate Gaia parallaxes 
indicating that they might be at a distance similar to $\eta$ Aql.  These 
cool stars are overwhelmingly 
not X-ray sources, that is not young enough to be associated with the Cepheid.  In this way 
 X-ray observations are valuable in sorting out possible structures in a crowded field in the 
galactic plane.

\section{Acknowledgements}

This research is 
based on observations obtained with XMM--Newton, an 
ESA science mission with instruments and contributions directly 
funded by ESA Member States and the USA (NASA).
 We thank the referee for comments that improved the presentation of 
the results.  
Support was provided to NRE by the Chandra X-ray Center NASA Contract NAS8-03060.
The observations were associated with program 84051
with support for this work from  
NASA Grant 80NSSC20K0794. JJD was supported by NASA contract NAS8-03060 to the 
 Chandra X-ray Center and thanks the Director, Pat Slane, for continuing advice and support.
HMG was supported through grant HST-GO-15861.005-A from the STScI under NASA contract NAS5-26555.

This work has made
use of data from the European Space Agency (ESA) mission Gaia (https://www.cosmos.esa.int/gaia), processed by
the Gaia Data Processing and Analysis Consortium (DPAC, https://www.cosmos.esa.int/web/gaia/dpac/consortium).
Funding for the DPAC has been provided by national institutions, in particular the institutions participating in the
Gaia Multilateral Agreement.
This publication makes use of data products from the Two Micron All Sky Survey, which is a joint project of 
the University of Massachusetts and the Infrared Processing and Analysis Center/California Institute of 
Technology, funded by the National Aeronautics and Space Administration and the National Science Foundation.

The SIMBAD database, and NASA’s Astrophysics Data System Bibliographic Services
were used in the preparation of this paper.

\hfil\vfill
\eject

\section{Appendix A: X-Ray Sources}

The full list of detected X-ray sources is provided here.
 The columns show Id, RA, Dec, position error, distance off axis, 
source significance, count rate and error, and exposure time.

\begin{deluxetable}{llccccccc}
\tablecaption{{\it XMM-Newton} Sources in the Image around $\eta$ Aql  \qquad\qquad \label{src}}
\tablewidth{0pt}
\tablehead{
 \colhead{} &  \colhead{RA}  & \colhead{Dec}  &  \colhead{Pos Err}  & \colhead{Offaxis}  & \colhead{Significance}
 &  \colhead{Rate}  & \colhead{$\pm$}  & \colhead{Exp Time}    \\
\colhead{} &  \colhead{(J2000) }  & \colhead{(J2000)} &  \colhead{ $"$ }
 & \colhead{$'$}  & \colhead{$\sigma_{bkg}$}
 &  \colhead{counts ks$^{-1}$}  & \colhead{counts ks$^{-1}$}  & \colhead{ks}    \\
}
\startdata
1 & 298.13559 & 0.77763 & 1.2 & 14.89 & 11.6 & 1 & 0.16 & 54.14   \\
2 & 298.15073 & 0.79184 & 2.9 & 14.23 & 5.5 & 0.61 & 0.15 & 58.38 \\
3 & 298.14593 & 0.79717 & 1.8 & 13.86 & 6.8 & 0.62 & 0.12 & 60.28 \\ 
4 & 298.17638 & 0.81929 & 4.8 & 13.13 & 7.3 & 1.3 & 0.21 & 66.91 \\
5 & 298.02853 & 0.82115 & 2.6 & 12.72 & 35.8 & 9.3 & 0.45 & 73.99 \\
6 & 298.12188 & 0.82154 & 2.9 & 12.16 & 24.6 & 5.7 & 0.35 & 69.57 \\
7 & 297.97294 & 0.83689 & 3.1 & 13.33 & 10.5 & 1.8 & 0.26 & 53.65 \\
8 & 298.22683 & 0.84742 & 3.2 & 13.14 & 8.1 & 1.1 & 0.16 & 70.23 \\
9 & 298.14715 & 0.8476 & 5.1 & 10.94 & 6.4 & 0.91 & 0.17 & 80.53 \\
10 & 297.9432 & 0.86631 & 3.6 & 13.07 & 7.2 & 1.1 & 0.2 & 66.54 \\
11 & 298.23292 & 0.86869 & 4.6 & 12.39 & 7.9 & 1.2 & 0.19 & 76.88 \\
12 & 298.06268 & 0.87544 & 1.9 & 9.03 & 16.3 & 1.3 & 0.14 & 107.1 \\
13 & 297.96099 & 0.88244 & 4.3 & 11.63 & 9.9 & 1.8 & 0.24 & 76.06 \\
14 & 297.91282 & 0.89002 & 2.8 & 13.52 & 12.1 & 2.5 & 0.3 & 54.42 \\
15 & 298.0837 & 0.90181 & 3.6 & 7.27 & 7.3 & 0.52 & 0.099 & 123.8  \\
16 & 297.96774 & 0.90337 & 5.1 & 10.46 & 6.6 & 1 & 0.19 & 85.73  \\
17 & 298.02259 & 0.90362 & 3.5 & 8.35 & 7.6 & 0.68 & 0.12 & 108.06 \\
18 & 298.05123 & 0.90412 & 3.1 & 7.57 & 10.8 & 0.89 & 0.12 & 120.05 \\
19 & 298.30902 & 0.90656 & 1.9 & 14.59 & 7.6 & 0.6 & 0.12 & 65.45 \\
20 & 297.90717 & 0.91499 & 2.6 & 12.99 & 8 & 0.84 & 0.16 & 61.47 \\
21 & 298.02152 & 0.91517 & 2.5 & 7.81 & 8 & 0.5 & 0.093 & 116.24 \\
22 & 298.06334 & 0.91631 & 2.4 & 6.65 & 18.8 & 1.6 & 0.14 & 136.84 \\
23 & 298.02735 & 0.9238 & 2.3 & 7.18 & 10.4 & 0.68 & 0.1 & 123.1 \\
24 & 298.0633 & 0.9264 & 1.8 & 6.07 & 37.4 & 4 & 0.2 & 144.38 \\
25 & 298.11144 & 0.92732 & 2.3 & 5.79 & 13 & 1.1 & 0.18 & 79.64 \\
26 & 297.98575 & 0.92903 & 1.2 & 8.63 & 11.6 & 0.58 & 0.095 & 96.88 \\
27 & 298.24136 & 0.93095 & 4.1 & 10.35 & 10 & 1.4 & 0.17 & 98.02 \\
28 & 298.1003 & 0.93627 & 4.1 & 5.18 & 5 & 0.29 & 0.071 & 164.33 \\
29 & 297.92155 & 0.93742 & 4.8 & 11.6 & 7.8 & 1.4 & 0.23 & 71.81 \\
30 & 298.24502 & 0.94465 & 4.1 & 10.13 & 11 & 1.2 & 0.15 & 125.83 \\
31 & 298.04614 & 0.95117 & 4.1 & 5.19 & 10.7 & 1.1 & 0.13 & 140.5 \\
32 & 298.02082 & 0.95377 & 4.2 & 6.07 & 5.6 & 0.45 & 0.11 & 104.81 \\
33 & 298.09953 & 0.95591 & 2.1 & 4 & 14.4 & 0.78 & 0.098 & 149.24 \\
34 & 298.18953 & 0.95716 & 1.6 & 6.88 & 53.8 & 5.8 & 0.23 & 166.02 \\
35 & 297.98522 & 0.9591 & 2.2 & 7.61 & 24.3 & 2.8 & 0.2 & 110.65 \\
36 & 298.00742 & 0.96017 & 1.5 & 6.46 & 59 & 9.1 & 0.32 & 128.11 \\
37 & 298.04064 & 0.96077 & 5.4 & 4.94 & 6.5 & 0.59 & 0.13 & 120.75 \\
38 & 298.08194 & 0.9616 & 1 & 3.74 & 177.8 & 39 & 0.55 & 173.23 \\
39 & 298.13599 & 0.96688 & 2.9 & 4.14 & 6.2 & 0.22 & 0.052 & 187.09 \\
40 & 298.11306 & 0.96833 & 2.8 & 3.42 & 13.8 & 0.79 & 0.09 & 186.28 \\
41 & 298.13129 & 0.96925 & 2.1 & 3.86 & 6.3 & 0.17 & 0.042 & 188.23 \\
42 & 298.15016 & 0.97458 & 1.8 & 4.37 & 18.2 & 0.88 & 0.087 & 189.34 \\
43 & 297.99777 & 0.97467 & 3.1 & 6.51 & 9.6 & 0.84 & 0.12 & 126.7 \\
44 & 298.0847 & 0.97541 & 3 & 2.89 & 5.1 & 0.22 & 0.057 & 181.51 \\
45 & 298.06138 & 0.97993 & 2.5 & 3.26 & 17.9 & 1.2 & 0.11 & 172.11 \\
46 & 298.17913 & 0.98355 & 2.7 & 5.55 & 6.8 & 0.28 & 0.058 & 182.57 \\
47 & 298.17005 & 0.98532 & 2.4 & 5.01 & 27.7 & 3 & 0.26 & 113.19 \\
48 & 298.22294 & 0.98806 & 5.5 & 7.94 & 5.2 & 0.49 & 0.11 & 128.83 \\
49 & 298.25923 & 0.99148 & 2.7 & 10.02 & 6.9 & 0.35 & 0.072 & 133.01 \\
50 & 298.18761 & 0.99902 & 3.7 & 5.72 & 6.1 & 0.34 & 0.068 & 182.67 \\
51 & 298.04709 & 1.00002 & 2.6 & 3.18 & 16 & 1 & 0.1 & 168.09 \\
52 & 297.94459 & 1.00089 & 1.5 & 9.13 & 6.5 & 0.29 & 0.074 & 94.95 \\
53 & 298.05841 & 1.00121 & 2.7 & 2.55 & 6.5 & 0.25 & 0.054 & 176.42 \\
54 & 298.15388 & 1.00602 & 1.9 & 3.66 & 20.6 & 1.6 & 0.18 & 104.92 \\
55 & 298.09711 & 1.00735 & 2.8 & 0.91 & 5.8 & 0.21 & 0.049 & 199.41 \\
56 & 298.02721 & 1.00878 & 3.7 & 4.16 & 6.3 & 0.4 & 0.078 & 154.36 \\
57 & 298.18464 & 1.00933 & 5.2 & 5.42 & 7 & 0.45 & 0.088 & 184.58 \\
58 & 298.09174 & 1.01073 & 2.5 & 0.73 & 18.5 & 0.98 & 0.093 & 197.09 \\
59 & 298.1055 & 1.01208 & 3.1 & 0.88 & 10.5 & 0.53 & 0.073 & 203.11 \\
60 & 298.13017 & 1.01675 & 2.3 & 2.13 & 9.8 & 0.38 & 0.058 & 206.54 \\
61 & 298.13874 & 1.02167 & 2.6 & 2.61 & 7.4 & 0.26 & 0.049 & 205.89 \\
62 & 298.10665 & 1.02253 & 2 & 0.69 & 6.3 & 0.16 & 0.038 & 203.19 \\
63 & 298.1281 & 1.02271 & 2.3 & 1.98 & 10.3 & 0.38 & 0.058 & 206.58 \\
64 & 298.27584 & 1.02633 & 4.2 & 10.84 & 5.1 & 0.4 & 0.098 & 118.83 \\
65 & 298.17799 & 1.02696 & 4.8 & 4.98 & 7.6 & 0.53 & 0.089 & 185.91 \\
66 & 297.99493 & 1.0319 & 3.8 & 6.04 & 7.5 & 0.88 & 0.18 & 73.49 \\
67 & 298.11918 & 1.03384 & 3.7 & 1.59 & 6.7 & 0.3 & 0.062 & 198.8 \\
68 & 298.24482 & 1.03522 & 3.1 & 9.01 & 13.8 & 1.3 & 0.17 & 112.93 \\
69 & 298.3522 & 1.03587 & 8.1 & 15.44 & 6 & 1.8 & 0.77 & 27.57 \\
70 & 298.05257 & 1.04793 & 2.4 & 2.98 & 9.3 & 0.38 & 0.063 & 169 \\
71 & 298.08637 & 1.04831 & 1.9 & 1.64 & 15.4 & 0.71 & 0.077 & 188.78 \\
72 & 298.19448 & 1.05006 & 1.5 & 6.18 & 6.6 & 0.15 & 0.039 & 175.68 \\
73 & 298.11757 & 1.05108 & 2.3 & 2.18 & 22 & 1.3 & 0.1 & 196.53 \\
74 & 298.18142 & 1.05829 & 3 & 5.6 & 10.4 & 0.61 & 0.082 & 179.84 \\
75 & 297.95056 & 1.0619 & 4 & 8.99 & 5.8 & 0.51 & 0.12 & 97.54 \\
76 & 298.10981 & 1.06309 & 2.4 & 2.59 & 18.5 & 1.1 & 0.099 & 189.29 \\
77 & 298.14678 & 1.06695 & 2.5 & 4.09 & 17.5 & 1.1 & 0.1 & 182.49 \\
78 & 298.10436 & 1.06798 & 3.1 & 2.79 & 4.9 & 0.18 & 0.049 & 178.82 \\
79 & 298.34609 & 1.0687 & 3.3 & 15.31 & 4.9 & 0.74 & 0.22 & 29.35 \\ 
80 & 298.03616 & 1.0688 & 3.7 & 4.5 & 6.7 & 0.42 & 0.081 & 150.6 \\
81 & 298.02734 & 1.07298 & 2.3 & 5.07 & 20.8 & 2.2 & 0.18 & 105.53 \\
82 & 297.98902 & 1.07629 & 3.5 & 7.14 & 16.9 & 2.1 & 0.19 & 116.67 \\
83 & 298.18087 & 1.08055 & 1.9 & 6.21 & 7.6 & 0.25 & 0.05 & 169.89 \\
84 & 297.97492 & 1.08134 & 5 & 8.03 & 7.6 & 0.94 & 0.17 & 94.26 \\
85 & 298.28372 & 1.08918 & 5.4 & 12 & 5.7 & 0.74 & 0.15 & 108.43 \\
86 & 298.26744 & 1.09018 & 2 & 11.11 & 32.3 & 4 & 0.23 & 117.23 \\
87 & 298.00354 & 1.09121 & 3 & 6.87 & 10.9 & 1 & 0.14 & 96.51 \\
88 & 298.04294 & 1.09202 & 5.5 & 5.22 & 5.3 & 0.53 & 0.11 & 115.35 \\
89 & 298.06952 & 1.09481 & 3.3 & 4.61 & 8.1 & 0.56 & 0.092 & 125.2 \\
90 & 298.1001 & 1.09626 & 2.6 & 4.44 & 16.6 & 1.1 & 0.11 & 157.93 \\
91 & 298.01647 & 1.0992 & 4 & 6.59 & 11 & 1.3 & 0.16 & 100.17 \\
92 & 298.25408 & 1.10147 & 1.6 & 10.65 & 5 & 0.16 & 0.046 & 120.52 \\
93 & 298.13129 & 1.10356 & 1.9 & 5.33 & 5.2 & 0.19 & 0.07 & 98.99 \\
94 & 298.15782 & 1.10747 & 2.8 & 6.34 & 12.8 & 0.87 & 0.099 & 157.22 \\
95 & 298.22525 & 1.10841 & 2.3 & 9.35 & 20.8 & 2 & 0.16 & 133.26 \\
96 & 298.19441 & 1.11164 & 3.1 & 8.01 & 10.8 & 0.77 & 0.11 & 133.34 \\
97 & 298.14093 & 1.11534 & 1.9 & 6.21 & 7.3 & 0.33 & 0.071 & 102.13 \\
98 & 298.05349 & 1.11651 & 2.1 & 6.17 & 6.2 & 0.26 & 0.064 & 110.04 \\
99 & 298.14173 & 1.12134 & 3.7 & 6.56 & 6.8 & 0.47 & 0.093 & 121.93 \\
100 & 297.93996 & 1.12587 & 5.4 & 11.19 & 5.8 & 0.97 & 0.2 & 79.17 \\
101 & 298.3107 & 1.12661 & 1.8 & 14.36 & 50 & 21 & 1 & 38.43 \\
102 & 298.26009 & 1.12883 & 2.9 & 11.77 & 5.8 & 0.38 & 0.09 & 108.4 \\
103 & 298.13912 & 1.1328 & 2.2 & 7.13 & 25.1 & 2.5 & 0.18 & 114.93 \\
104 & 298.17603 & 1.13296 & 4 & 8.22 & 5.7 & 0.37 & 0.087 & 137.52 \\
105 & 298.20407 & 1.13579 & 5.3 & 9.43 & 8.4 & 0.72 & 0.15 & 112.3 \\
106 & 298.16633 & 1.13661 & 2.9 & 8.07 & 11.3 & 0.97 & 0.12 & 109.35 \\
107 & 298.30523 & 1.13966 & 5.3 & 14.43 & 8.1 & 2 & 0.41 & 38.05 \\
108 & 298.2856 & 1.14716 & 4.4 & 13.66 & 5.2 & 0.89 & 0.24 & 37.87 \\
109 & 297.95861 & 1.14855 & 2.6 & 11.15 & 15.9 & 2.8 & 0.27 & 60.68 \\
110 & 297.9167 & 1.1491 & 5.4 & 13.13 & 3153 & 58000 & 38000 & 33.24 \\
111 & 298.04519 & 1.15281 & 2.9 & 8.38 & 5.3 & 0.37 & 0.086 & 89.51 \\
112 & 298.12494 & 1.1562 & 3.1 & 8.22 & 9.2 & 0.88 & 0.13 & 101.4 \\
113 & 298.28088 & 1.16909 & 2.2 & 14.2 & 4.8 & 0.55 & 0.28 & 18.13 \\
114 & 298.26034 & 1.16911 & 7.1 & 13.25 & 4.9 & 1.5 & 0.52 & 32.89 \\
115 & 298.02206 & 1.17155 & 3.9 & 9.96 & 11.6 & 2.2 & 0.26 & 68.2 \\
116 & 298.13075 & 1.17726 & 5.1 & 9.53 & 5.9 & 0.77 & 0.15 & 90.34 \\
117 & 298.19823 & 1.18183 & 3.6 & 11.39 & 15.3 & 2.1 & 0.2 & 104.46 \\
118 & 298.08906 & 1.18365 & 2.3 & 9.68 & 20.8 & 2.7 & 0.21 & 85.57 \\
119 & 298.11909 & 1.18782 & 4.4 & 10.03 & 9 & 1.3 & 0.18 & 85.62 \\
120 & 298.09918 & 1.19779 & 2 & 10.52 & 6.2 & 0.42 & 0.098 & 72.35 \\
121 & 298.22869 & 1.20235 & 2.9 & 13.44 & 16.1 & 3.4 & 0.43 & 38.87 \\
122 & 298.21207 & 1.21589 & 3.5 & 13.56 & 5 & 0.65 & 0.21 & 37.97 \\
123 & 298.17517 & 1.21755 & 3.9 & 12.65 & 6.4 & 0.73 & 0.16 & 76.13 \\
124 & 298.03944 & 1.22004 & 4.8 & 12.32 & 9.1 & 3.2 & 1 & 16.8 \\
125 & 298.12977 & 1.23496 & 1.6 & 12.92 & 7.4 & 1 & 0.28 & 19.37 \\
126 & 298.10656 & 0.7609 & 5.3 & 15.71 & 102.1 & 76 & 48 & 31.03 \\
\enddata

\end{deluxetable}

\hfil\vfill
\eject

\section{Appendix B: Cross matches with 2MASS sources}
The full list of 2MASS sources which coincide with XMM-Newton sources is provided here, 
 The columns show the XMM-Newton source ID, the RA, the Dec, the 
2MASS ID, the 2MASS magnitudes and errors (J, H, and K), the 2MASS quality indicators, 
and the separation between the X-ray source and the 2MASS source.




\begin{deluxetable}{llcccccccccc}
\tablecaption{2MASS Counterparts to   {\it XMM-Newton} Sources   \qquad\qquad \label{src}}
\tablewidth{0pt}
\tablehead{
 \colhead{} &  \colhead{RA}  & \colhead{Dec}  &  \colhead{2MASS ID}  & \colhead{J}  & \colhead{$\pm$}
 &  \colhead{H}  & \colhead{$\pm$}  & \colhead{K}   & \colhead{$\pm$} & \colhead{Qual}  & \colhead{Sep}   \\
\colhead{} &  \colhead{(J2000) }  & \colhead{(J2000)} &  \colhead{  }
 & \colhead{mag}  & \colhead{mag}
 &  \colhead{mag}  & \colhead{mag}  & \colhead{mag}  & \colhead{mag} & \colhead{}  
& \colhead{$"$}   \\
}
\startdata
 11.0 & 298.233307 & 0.868617 &  19525599+0052070 &  15.7430 &  0.0910 &  15.3750 &  0.1150 &  15.1300 &  0.1450 & ABB &  1.3929 \\
 18.0 & 298.051544  & 0.903709 &  19521236+0054133 &  14.0700 &  0.0780 &  13.5690 &  0.1130 &  13.5390 &  0.0580 & ABA &  1.0000 \\
 18.0 & 298.052582 & 0.903914 &  19521261+0054140 &  14.7210 &  0.0450 &  14.0680 &  0.0580 &  14.1160 &  0.0610 & AAA &  1.0000 \\
 21.0 &   298.021759 & 0.914922 &  19520521+0054537 &  14.3140 &  0.0290 &  13.7460 &  0.0410 &  13.6970 &  0.0360 & AAA &  1.2030 \\
 23.0  &  298.027588 & 0.923320 &  19520662+0055239 &  12.4940 &  0.0240 &  11.8010 &  0.0230 &  11.6570 &  0.0230 & AAA &  1.9533 \\
 25.0 &   298.111786 & 0.926176 &  19522682+0055342 &  16.0190 &  0.0950 &  15.4560 &  0.1140 &  15.3290 &  0.1690 & ABC &  4.2903 \\
 26.0  &  297.984619 & 0.929073 &  19515630+0055446 &  13.7680 &  0.0300 &  13.4280 &  0.0380 &  13.2970 &  0.0330 & AAA &  4.0812 \\
 27.0 &   298.240326 & 0.930857 &  19525768+0055510 &  15.9650 &  0.1070 &  15.2150 &  0.1300 &  14.7140 &  0.1110 & ABB &  3.7047 \\
 28.0 &   298.100220 & 0.936506 &  19522405+0056114 &  16.2620 &  0.1170 &  15.5780 &  0.1250 &  15.5910 &   NaN & BBU &  0.9080 \\
 31.0 &   298.046539 & 0.951528 &  19521117+0057055 &  15.0410 &  0.0430 &  14.2810 &  0.0610 &  14.1480 &  0.0700 & AAA &  1.9431 \\
 34.0  &   298.189575 & 0.956945 &  19524549+0057250 &  12.9770 &  0.0220 &  12.2930 &  0.0250 &  12.1520 &  0.0260 & AAA &  0.7848 \\
 35.0 &   297.984985 & 0.959162 &  19515639+0057329 &  12.9000 &  0.0270 &  12.2990 &  0.0330 &  12.0580 &  0.0270 & AAA &  0.8748 \\
 38.0 &    298.082062 & 0.961234 &  19521969+0057404 &  7.3540 &  0.0270 &  6.8600 &  0.0270 &  6.6760 &  0.0180 & AAA &  1.3738 \\
 39.0 &   298.136780 & 0.966529 &  19523282+0057595 &  13.5340 &  0.0510 &  13.2020 &  0.0730 &  13.1460 &  0.0670 & AAA &  3.1052 \\
 40.0 &   298.113007 & 0.968156 &  19522712+0058053 &  13.5230 &  0.0260 &  12.7800 &  0.0290 &  12.5280 &  0.0240 & AAA &  0.6420 \\
 44.0  &   298.084045 & 0.975207 &  19522017+0058307 &  15.9070 &  0.0800 &  15.5380 &  0.1250 &  15.2580 &  0.1490 & ABC &  2.4477 \\
 52.0 &   297.944611 & 1.000383 &  19514670+0100013 &  12.7120 &  0.0220 &  12.5270 &  0.0210 &  12.4570 &  0.0230 & AAA &  1.8270 \\
 53.0 &   298.058807 & 1.000881 &  19521411+0100031 &  16.7820 &  0.1530 &  15.9200 &  0.1880 &  15.8040 &  0.2420 & CCD &  1.8782 \\
 62.0  &  298.106659 & 1.022538 &  19522560+0101211 &  14.4550 &    NaN &  14.7810 &  0.0710 &  14.6430 &  0.0950 & UAA &  0.0677 \\
 67.0 &   298.119720 & 1.034258 &  19522873+0102033 &  16.3270 &  0.1540 &  15.7630 &  0.1820 &  15.3190 &  0.1860 & BCC &  2.4811 \\
 74.0 &   298.180420 & 1.057761 &  19524330+0103279 &  15.2100 &  0.0510 &  14.9150 &  0.0800 &  14.8950 &  0.1220 & AAB &  4.0340 \\
 78.0 &   298.103333 & 1.068579 &  19522479+0104068 &  15.9470 &  0.1090 &  15.2790 &  0.0990 &  15.7170 &  0.2440 & AAD &  4.3262 \\
 80.0 &   298.036011 & 1.069712 &  19520864+0104109 &  15.0910 &  0.0340 &  14.5500 &  0.0620 &  14.5100 &  0.0780 & AAA &  3.3191 \\
 82.0  &   297.989624 & 1.076165 &  19515750+0104341 &  14.3100 &  0.0520 &  13.5240 &  0.0530 &  13.2610 &  0.0400 & AAA &  2.0000 \\
 82.0 &   297.988983 & 1.075198 &  19515735+0104307 &  15.8580 &  0.0960 &  15.0530 &  0.1100 &  14.7330 &  0.0970 & ABA &  2.0000 \\
 87.0 &   298.003479 & 1.091180 &  19520083+0105282 &  14.1140 &  0.0380 &  13.4370 &  0.0440 &  13.2450 &  0.0350 & AAA &  0.2447 \\
 90.0  &  298.099823 & 1.095481 &  19522395+0105437 &  16.4430 &  0.1490 &  16.1620 &  0.2230 &  15.7880 &    NaN & BDU &  2.9774 \\
 91.0 &   298.016022 & 1.099517 &  19520384+0105582 &  14.4800 &  0.0340 &  13.7790 &  0.0420 &  13.6940 &  0.0440 & AAA &  1.9813 \\
 93.0 &   298.131683 & 1.103213 &  19523160+0106115 &  14.8900 &  0.0430 &  14.2740 &  0.0540 &  13.9760 &  0.0680 & AAA &  1.8576 \\
 94.0 &   298.158081 & 1.107010 &  19523793+0106252 &  13.9660 &  0.0310 &  13.3720 &  0.0340 &  13.0810 &  0.0350 & AAA &  1.8917 \\
 95.0 &   298.225220 & 1.107712 &  19525405+0106277 &  15.5760 &  0.0720 &  14.9040 &  0.0790 &  14.8640 &  0.1200 & AAB &  2.5167 \\
 96.0 &   298.195404 & 1.112088 &  19524689+0106435 &  15.8310 &  0.0900 &  15.2700 &  0.1120 &  14.7550 &  0.1090 & ABB &  3.9376 \\
101.0 &   298.310669 & 1.126410 &  19531455+0107350 &  10.8820 &  0.0240 &  10.2870 &  0.0330 &  10.0790 &  0.0210 & AAA &  0.7330 \\
109.0 &   297.958771 & 1.148864 &  19515010+0108559 &  15.4870 &  0.0920 &  14.6480 &  0.0890 &  14.1650 &  0.0780 & AAA &  1.2495 \\
113.0  &  298.280731 & 1.167832 &  19530737+0110041 &  16.4870 &  0.1560 &  15.7260 &  0.1700 &  15.3100 &  0.1720 & BCC &  4.5640 \\
120.0 &   298.099304 & 1.197495 &  19522383+0111509 &  13.9210 &  0.0340 &  13.5720 &  0.0300 &  13.3540 &  0.0420 & AAA &  1.1439 \\
125.0 &   298.130981 & 1.235249 &  19523143+0114068 &  15.3210 &  0.0680 &  14.8910 &  0.0860 &  14.5800 &  0.1000 & AAA &  4.5301 \\
\enddata
\end{deluxetable}



\section{Appendix C: Cross Match with {\it Gaia} Sources}
The list of Gaia sources that coincide with X-ray sources is provided here. The colunms 
The columns show the RA, the Dec, the separation of the XMM-Newton source and the Gaia source, the Gaia ID, the
XMM-Newton source number, the parallax and error, and the Gaia magnitudes and errors (G, BP and RP).

\hfil\vfill
\eject

\begin{table*}
\caption{ Gaia Counterparts to XMM-Newton Sources   \qquad\qquad \label{src}}
\resizebox{0.99\textwidth}{!}{
  \begin{tabular}{llccccccccccc}\hline\hline
\tablewidth{0pt}
RA  & Dec  &  Sep  & Gaia ID & XMM-Newton Src \# & Px & ePx &  G  & eG  & BP  & eBP & RP & eRP  \\
 $^o$  & $^o$  & $"$  &          &          & mas & mas  &  mag  & mag  & mag   & mag & mag  & mag   \\
\\hline
298.17575892226 & 0.81915965345 & 0.45  & 4240251955274546176  & 4   & -0.2996 & 0.3579 & 19.3837 & 0.0036 & 19.6484 & 0.0707 & 18.7031 & 0.0462 \\
298.02863373075 & 0.82086525281 & 0.3  & 4240262920333547008  & 5   & 0.5077 & 1.1782 & 20.6907 & 0.0097 & 20.5658 & 0.1153 & 20.0888 & 0.1044 \\
298.12175960105 & 0.82134431695 & 0.22  & 4240252470670622720  & 6   & 0.7396 & 1.0209 & 20.5027 & 0.0093 & 20.471 & 0.1214 & 19.4289 & 0.0448 \\
297.97295268783 & 0.83683043657 & 0.06  & 4240263122188744704  & 7   & 0.1319 & 0.6752 & 20.1036 & 0.0095 & 20.3039 & 0.0777 & 19.5392 & 0.0558 \\
297.94457964325 & 0.86643868687 & 0.96  & 4240264260358061056  & 10 & 1.0212 & 1.6097 & 20.6207 & 0.0127 & 19.7848 & 0.1521 & 18.8962 & 0.0585 \\
297.94414349252 & 0.8669634727  & 0.81  & 4240264260363349504  & 10 &  -1.659 & 1.3652 & 20.6232 & 0.0117 & 20.9315 & 0.1141 & 19.8612 & 0.1273 \\
298.2324697473 & 0.86799299322 & 0.62  & 4240258041244867072  & 11 & 0.1434 & 0.187 & 18.2923 & 0.0019 & 18.8127 & 0.0178 & 17.6504 & 0.0135 \\
298.23333612618 & 0.86857999723 & 0.33  & 4240258041250142720  & 11 & 0.6379 & 0.1283 & 17.7589 & 0.0026 & 18.4189 & 0.0159 & 16.9528 & 0.0081 \\
298.06322509666 & 0.87504033519 & 0.91  & 4240265011974489088  & 12 &  0.0236 & 0.7892 & 20.3472 & 0.0084 & 20.2782 & 0.0853 & 19.7389 & 0.0588 \\
298.08428431342 & 0.90171111212 & 0.54  & 4240265355571958784  & 15 & 0.8226 & 0.3363 & 19.2625 & 0.003 & 19.854 & 0.0746 & 18.288 & 0.0203 \\
297.96899077301 & 0.90355485948 & 0.84  & 4240265875265839616  & 16 & 1.6979 & 0.4308 & 19.6082 & 0.0038 & 20.2322 & 0.1392 & 18.5524 & 0.0396 \\
298.05127801538 & 0.90382990189 & 0.32  & 4240265325509828096  & 18 & 0.5702 & 0.1587 & 17.5987 & 0.0112 & NA &  NA & NA & NA \\
298.05162887172 & 0.90368197187 & 0.69  & 4240265325514908160  & 18 & 0.635 & 0.0526 & 15.4858 & 8e-04 & 15.8692 & 0.0067 & 14.7899 & 0.0099 \\
298.02170774622 & 0.91493841657 & 0.42  & 4240266115788933632  & 21 & 0.3413 & 0.0555 & 16.1135 & 0.0015 & 16.7264 & 0.0094 & 15.3481 & 0.007 \\
298.02759556192 & 0.92331478358 & 0.85  & 4240266150148679168  & 23 & 0.1726 & 0.0279 & 14.4374 & 0.0016 & 15.1793 & 0.0088 & 13.6003 & 0.0048 \\
298.10023053976 & 0.93650971902 & 0.21  & 4240271368526282752  & 28 & 0.662 & 0.1625 & 17.9773 & 0.0019 & 18.3262 & 0.0521 & 17.1773 & 0.0101 \\
297.92141916554 & 0.93662542908 & 0.58  & 4240269139446042624  & 29 & -0.0493 & 0.2696 & 18.9251 & 0.0027 & 19.3926 & 0.0229 & 18.3257 & 0.0239 \\
298.24511145377 & 0.94352047395 & 0.92  & 4240261167981228544  & 30 & 0.2451 & 0.3457 & 19.2372 & 0.0043 & NA &  NA &  NA & NA \\
298.04651516881 & 0.95147852453 & 0.42  & 4240266493746085888  & 31 & 0.8268 & 0.0873 & 16.985 & 0.0032 & 17.6866 & 0.015 & 16.1351 & 0.0092 \\
298.18954120715 & 0.95689654593 & 0.58  & 4240271682066449408  & 34 & 2.1357 & 0.0359 & 14.9791 & 0.0015 & 15.7388 & 0.0055 & 14.1066 & 0.0054 \\
297.98509285273 & 0.959210677  & 0.27  & 4240269448683655680  & 35 & 10.7134 & 0.0757 & 15.9245 & 7e-04 & 17.5694 & 0.008 & 14.677 & 0.0017 \\
298.11303036188 & 0.96805563443 & 0.35  & 4240272266182088704  & 40 & 2.4875 & 0.0739 & 16.3465 & 0.0012 & 17.4999 & 0.0178 & 15.152 & 0.0035 \\
298.08404267921 & 0.97525372621 & 0.78  & 4240272506700296704  & 44 & 0.2905 & 0.1025 & 17.3575 & 0.0012 & 17.678 & 0.0214 & 16.7159 & 0.0084 \\
298.08382850926 & 0.97475575375 & 0.92  & 4240272506695313920  & 44 & 1.0223 & 1.2996 & 20.3331 & 0.0133 & NA &  NA & NA & NA \\
298.22222691217 & 0.98739569896 & 0.58  & 4240273223952942080  & 48 & 0.3885 & 0.6336 & 19.8115 & 0.0046 & 20.074 & 0.0789 & 18.8681 & 0.0266 \\
298.25907408523 & 0.99230678489 & 0.89  & 4240261614657941504  & 49 & 0.5041 & 0.8449 & 20.1284 & 0.0055 & 20.2235 & 0.0704 & 19.2216 & 0.0422 \\
298.04803223586 & 1.00046627072 & 0.99  & 4240275599071934976  & 51 & 0.2742 & 1.143 & 20.6241 & 0.0098 & 20.7255 & 0.1452 & 19.2588 & 0.0517 \\
298.05868727163 & 1.00083276945 & 0.58  & 4240275530357335040  & 53 & 0.3288 & 0.1891 & 18.37 & 0.0031 & 18.7961 & 0.0224 & 17.5494 & 0.0199 \\
298.10580789371 & 1.01194128494 & 0.29  & 4240273022092038144  & 59 & -2.1311 & 1.5816 & 20.4022 & 0.0178 & NA & NA & NA & NA \\
298.10664606908 & 1.02255603921 & 0.04  & 4240273017795262464  & 62 & 1.888 & 0.2412 & 18.4717 & 0.0032 & 19.0428 & 0.0578 & 17.1374 & 0.0087 \\
298.11974050898 & 1.03434032988 & 0.7  & 4240273159535376384  & 67 & 0.1128 & 0.146 & 17.7889 & 0.0014 & 18.1199 & 0.0138 & 17.05 & 0.0056 \\
298.35346738951 & 1.0346965463  & 0.76  & 4240355760344077824  & 69 & 0.0998 & 0.19  & 17.4827 & 0.0018 & 17.921 & 0.0342 & 16.5814 & 0.0209 \\
298.35247414448 & 1.03748925609 & 0.7  & 4240355756040043008  & 69 & 0.216 & 0.5383 & 19.4351 & 0.0037 & 19.9638 & 0.0488 & 18.5775 & 0.0337 \\
298.03600315316 & 1.06969286488 & 0.87  & 4240277145265164288  & 80 &  0.5103 & 0.0748 & 16.4506 & 8e-04 & 16.8905 & 0.0072 & 15.815 & 0.003 \\
297.98963927238 & 1.07616154178 & 0.6  & 4240277007826251264  & 82 & 3.1203 & 0.4037 & 17.4586 & 0.0028 & 19.0471 & 0.0403 & 16.1247 & 0.0037 \\
297.97436262785 & 1.0816724037  & 0.43  & 4240464680707048448  & 84 &  0.8924 & 0.4434 & 19.5996 & 0.0043 & 19.9888 & 0.0404 & 18.8989 & 0.0294 \\
298.28486376792 & 1.08870916887 & 0.66  & 4240368125550486016  & 85 & 0.7317 & 1.5874 & 20.8131 & 0.0177 & 21.253 & 0.2294 & 20.3717 & 0.1491 \\
298.00346167643 & 1.09108679338 & 0.17  & 4240465058672031232  & 87 & 1.2554 & 0.0562 & 16.0268 & 0.0034 & 16.7202 & 0.0127 & 15.2213 & 0.0145 \\
298.0160327258 & 1.09950184906 & 0.47  & 4240277420143114240  & 91 & 0.3293 & 0.0682 & 16.3634 & 0.0043 & 17.0758 & 0.0152 & 15.5227 & 0.0148 \\
298.15806340676 & 1.10693701577 & 0.63  & 4240278206114537472  & 94 & 4.353 & 0.7044 & 17.6788 & 0.0088 & 18.8616 & 0.0269 & 15.8817 & 0.0029 \\
298.1411321078 & 1.11527381957 & 0.35  & 4240278481003275776  & 97 & 0.0307 & 0.31  & 18.91 & 0.0028 & 19.3476 & 0.0277 & 18.1713 & 0.017 \\
298.31069099153 & 1.12644699558 & 0.32  & 4240368645240608768  & 101  & 0.8175 & 0.0328 & 13.0266 & 0.0029 & 13.5312 & 0.015 & 12.0358 & 0.0069 \\
298.20439572941 & 1.13529175691 & 0.37  & 4240372214359375872  & 105  & 0.7619 & 0.6112 & 19.8206 & 0.0051 & 20.1968 & 0.0752 & 18.9805 & 0.0501 \\
298.16545949101 & 1.13645850149 & 0.82  & 4240278618431430144  & 106  & -1.5236 & 0.9903 & 20.4576 & 0.0084 & 20.4093 & 0.1259 & 19.534 & 0.0695 \\
298.12504192405 & 1.1568237908  & 0.51  & 4240279653521397248  & 112  & 0.09  & 1.4437 & 20.6067 & 0.0109 & 21.0408 & 0.1805 & 19.8127 & 0.0866 \\
298.25912778959 & 1.16764207983 & 0.94  & 4240372695396496896  & 114  & 0.6423 & 0.2298 & 18.2577 & 0.0018 & 18.5663 & 0.0401 & 17.2984 & 0.0377 \\
298.25953148388 & 1.16745112385 & 0.90  & 4240372695395740672  & 114  & 0.4905 & 0.3685 & 18.6138 & 0.0041 & NA &  NA & NA & NA \\
298.26010068621 & 1.16901821391 & 0.12  & 4240372699695164416  & 114  & 1.2905 & 0.627 & 19.8021 & 0.0055 & 20.163 & 0.0929 & 19.0334 & 0.0379 \\
298.11983168352 & 1.18843671759 & 0.73  & 4240279756600641536  & 119  & -0.1333 & 0.5125 & 19.4829 & 0.004 & 20.2056 & 0.0627 & 18.6513 & 0.0247 \\
298.09935425403 & 1.19746072366 & 0.65  & 4240467498213515264  & 120  & 0.3669 & 0.0535 & 15.6014 & 0.0015 & 16.1671 & 0.0057 & 14.8851 & 0.0056 \\
298.03975170968 & 1.21907311088 & 0.66  & 4240468455985499648  & 124  & 2.4464 & 0.8865 & 20.2308 & 0.0075 & 20.5786 & 0.0919 & 19.4227 & 0.0591  \\
298.03865207232 & 1.21953314898 & 0.65  & 4240468455984104960  & 124  & 0.2874 & 0.5161 & 19.6723 & 0.0042 & 20.1114 & 0.0701 & 19.0047 & 0.0319 \\
298.10528737867 & 0.76176863706 & 0.95  & 4240249790611020288  & 126  & -0.3336 & 0.6307 & 19.9983 & 0.0052 & 20.6011 & 0.0806 & 19.1697 & 0.0354 \\
\end{tabular}
}
\end{table*}






\end{document}